\documentclass[english,aps, prl,  twocolumn, superscriptaddress]{revtex4-1}
\usepackage[utf8]{inputenc}
\usepackage[T1]{fontenc}
\usepackage{color}
\usepackage{babel}
\usepackage{amsmath}
\usepackage{amssymb}
\usepackage{graphicx}
\usepackage{esint}
\usepackage{times}
\usepackage[unicode=true,pdfusetitle,
 bookmarks=true,bookmarksnumbered=false,bookmarksopen=false,
 breaklinks=false,pdfborder={0 0 1},backref=false,colorlinks=true,citecolor=blue,
 urlcolor=blue,linkcolor=blue]
 {hyperref}
\usepackage{breakurl}

\makeatletter
\usepackage{breqn}

\makeatletter
\let\cat@comma@active\@empty
\makeatother

\begin{document}

\title{Statistics of fractionalized excitations through threshold
spectroscopy}

\author{Siddhardh C. Morampudi}
\affiliation{Max-Planck-Institut f\"ur Physik komplexer Systeme, 01187 Dresden, Germany}
\author{Ari M. Turner}
\affiliation{Department of Physics and Astronomy, Johns Hopkins University, Baltimore, MD 21218, USA}
\author{Frank Pollmann}
\affiliation{Max-Planck-Institut f\"ur Physik komplexer Systeme, 01187 Dresden, Germany}
\author{Frank Wilczek}
\affiliation{Center for Theoretical Physics, MIT, Cambridge MA 02139, USA}
\affiliation{Department of Physics, Stockholm University, Stockholm Sweden}
\affiliation{Department of Physics and Origins Project, Arizona State University, Tempe AZ 25287 USA}
\affiliation{Wilczek Quantum Center, Zhejiang University of Technology, Hangzhou China}

\begin{abstract}
We show that the anyonic statistics of fractionalized excitations
display characteristic signatures in threshold spectroscopic measurements. Drawing motivation
from topologically ordered phases such as gapped quantum spin liquids and fractional chern insulators 
which possess fractionalized excitations, we consider gapped systems with abelian 
anyonic excitations. The low energy onset of associated correlation functions near the threshold show universal behaviour depending on the statistics of the anyons. This explains some recent 
theoretical results in spin systems and also provides a route 
towards detecting statistics in experiments such as neutron scattering and tunneling spectroscopy.

\end{abstract}

 \maketitle

Quantum mechanics allows for the possibility of phases that are not characterized by symmetry breaking, including the complex of the fractional quantum Hall effects\cite{Tsui1982,Nayak2008}. These are characterized from a theoretical point of view by subtle characteristics of their quantum entanglement known as topological order\cite{Wen1990,Wen2002,*Chen2010,*Essin2013,Kitaev2006,*Levin2006}.This phenomenon may also appear in quantum spin liquids, of which we have several candidates\cite{AndersonRVB1987,Balents2010,Moessner2001}.
From an experimental point of view the most interesting phenomenon in such phases is not the lack of long-range order -- a negative characteristic --  but rather the existence of fractionalized quasiparticle excitations that have anyonic statistics\cite{Leinaas1977,PhysRevLett.49.957,PhysRevLett.48.1144}.  In this paper we suggest a way to measure these quantum statistics through the threshold behavior
of spectroscopic cross-sections for creating these excitations.  This method is simpler for spin liquid materials than the interference phenomena that have been studied in the fractional quantum Hall effect.

Existing methods to determine those statistics are overwhelmingly theoretical.  They rely on
explicit microscopic calculations determining the Berry phase acquired
on exchange of particles in trial wavefunctions\cite{PhysRevB.35.8865, *PhysRevB.40.7133};
or on numerical methods, such as recent techniques utilizing degenerate ground states on a torus\cite{Zhang2012, *Moradi2015, *Zaletel2013, *Vidal2013, *Morampudi2014}. Proposals for
experimental signatures include proposals of performing
anyonic interferometry\cite{PhysRevB.72.075342, *Willett2009, *Bonderson2008} and various schemes to measure non-abelian statistics through the entropy associated with the quasiparticles\cite{Yang2009, *Cooper2009, *Yang2010, *Laumann2012, *Laumann2013}.  Clearly, it is desirable to identify additional signatures of braiding statistics in quantities accessible to established experimental techniques. 

Spectroscopic studies, such as scattering experiments, are a plausible place to seek such signatures. In classic work, Wigner demonstrated that the onset behavior of cross-sections near production thresholds is often dominated by long-range interactions (electric charge, or statistical ``interactions'' governing effective centrifugal forces)\cite{Wigner1948}.
Relevant cross-sections often can be related to dynamic correlation functions (spectral functions), such as the dynamic structure factor in inelastic neutron scattering studies of spin systems. This connection has already
been used to demonstrate signatures of fractionalization in candidate
spin liquids\cite{Han2012,Punk2014,PhysRevB.88.224413}.  Here the
corresponding characteristic feature is the absence of sharp dispersive features (e.g., magnons). They are replaced
by a broad continuum, plausibly interpreted as arising from different ways of sharing
momentum among the fractionalized quasiparticles, i.e., from their center of
mass motion.  It is natural, given such exotic fractional quasiparticles, to investigate whether they also exhibit fractional, "anyon" statistics, accumulating unusual phase factors as they braid around one another.  Such anyons have already been predicted theoretically in several candidate phases.
 
We will derive the behaviour of the cross-section near threshold for two free anyons analytically, and also perform numerics to 
verify the robustness of that behaviour to interactions.  We then discuss which types of excitations one could identify using this method experimentally in spin liquids. We will also use our method to rederive recent 
results for the dynamic structure factor in topologically ordered phases. Extending our results, we explore
the corresponding behavior for three particles.  This too displays an interesting dependence on the statistics, reflecting the braiding of three particles.

\begin{figure}
\includegraphics[width=1\columnwidth]{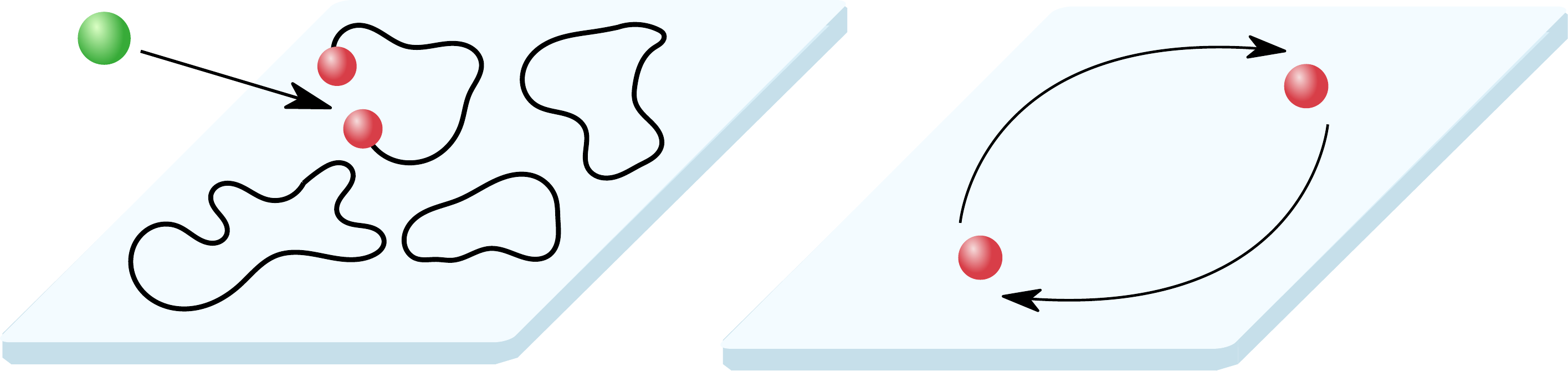}

\hspace*{-1.2cm}(a) ~~~~~~~~~~~~~~~~~~~~~~~~~~~~~~~~~~~~~~~~~~~~~(b)\protect\caption{(a) In phases with topological order, scattering processes can create fractionalized anyonic excitations
above some threshold energy.  (b) The statistics of the excitations mediate long-range effects between them\label{fig:neutron_anyon}}
\end{figure}

\paragraph*{Set-up and eigenvalue problem}
We consider a 2D system 
in which gapped fractionalized excitations can be created
by the action of a local operator (Fig.~\ref{fig:neutron_anyon}).
For example, we can consider inelastic neutron scattering on a gapped
spin liquid, resulting in the creation of
$n$ fractionalized excitations, which in general obey anyonic
exchange and mutual statistics. The double differential scattering cross-section
$d^{2}\sigma/d\Omega d\omega$ thus obtained is proportional to the
dynamic structure factor $S^{\alpha \beta}(\vec{q},\omega)$, defined as the fourier-transform of the correlation function $\langle S^{\alpha}(r, t) S^{\beta}(0, 0) \rangle$ where $\alpha, \beta = \{x, y, z\}$\cite{Zhu2005}. Since the excitations
are gapped, we can focus on the effective $n$-particle system to
calculate $S(\vec{q},\omega)$ at energies below those involving additional excitations.

An anyon can be viewed\cite{PhysRevLett.49.957}
as a composite particle of carrying ``electric'' charge and attached to an infinitesimally
thin ``magnetic'' flux tube (where the charge and flux pertain to an emergent gauge field, not ordinary electromagnetism). The
braiding phases arise as effective Aharonov-Bohm phases as these
composite particles are exchanged or taken around one another. We will
work in the magnetic/boson gauge, where the Hamiltonian acts on bosonic
wavefunctions and the statistics is encoded in the Hamiltonian as
an interaction through minimal coupling of an effective gauge field
$\vec{a}$. 

In the center of mass ($\vec{R})$ frame, 
the relative Hamiltonian depends on the statistics.
For the case of two identical fractionalized excitations\cite{Arovas1985}, $\vec{a}=\frac{\hbar c\alpha}{q}\nabla\theta$
where $\theta$ is the angle between the particles, $q$ is the charge and $\alpha$ is the statistics parameter which varies from $0$
for bosons to $1$ for fermions. Thus, for two excitations with a
quadratic dispersion, the effective Hamiltonian is given by

\begin{gather}
H_{R}=\frac{P_{\vec{R}}^{2}}{4m}\\
H_{r}=\frac{p_{r}^{2}}{m}+\frac{(p_{\phi}-\hbar\alpha)^{2}}{mr^{2}}+V\left(r,\phi\right)
\end{gather}

\noindent where $\vec{R}$ is the center of mass co-ordinate and $m,r,\phi$ are the mass and relative co-ordinates of the particles, and $V\left(r,\phi\right)$ is the effective interaction
between the two particles.

\begin{figure}

\includegraphics[width=0.9\columnwidth, keepaspectratio=True]{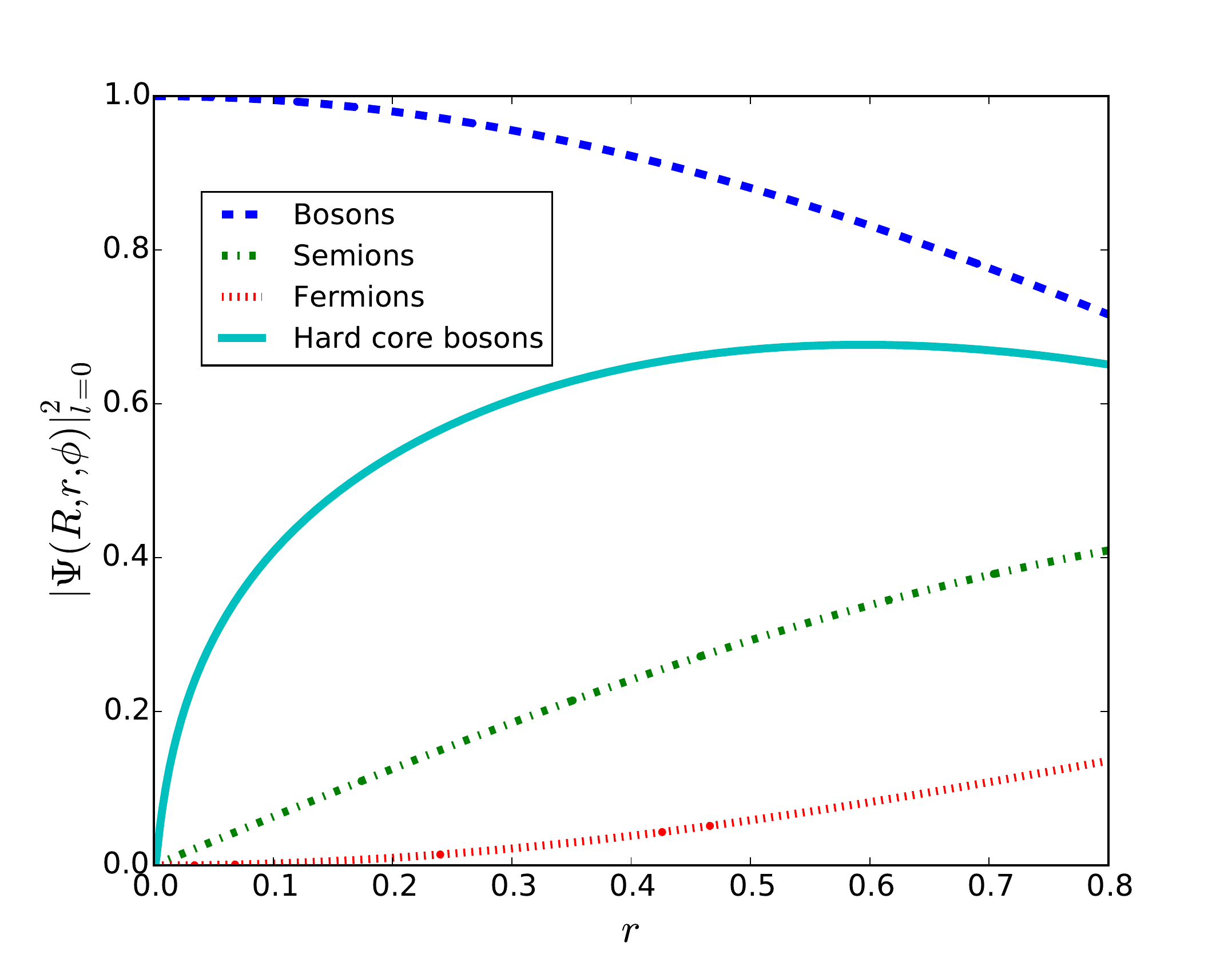}\protect\caption{Probability densities for various two identical anyon systems as a function
of their relative distance $r$ at a fixed $k$. The s-wave eigenstates ($l=0$) will
dominate the low-energy behaviour of the spectral function.\label{fig:wavefunction}}

\end{figure}

We temporarily set $V\left(r,\phi\right)=0$. Then we can solve the
eigenvalue problem for $H_{r}$ using separation of variables.
The complete normalized solution to the eigenvalue problem for $H=H_{R}+H_{r}$
is 

\begin{equation}
\Psi(\vec{R},r,\phi)\sim\sqrt{\dfrac{k}{L^3}}J_{|l-\alpha|}\left(kr\right)\exp(il\phi)\exp(i\vec{K}\cdot\vec{R})
\end{equation}

\noindent where $l=2n,\,n\in\mathbb{Z}$; relative momentum $k =\sqrt{\frac{E_{r}m}{\hbar^{2}}}$, 
center of mass momentum $\left|\vec{K}\right|=2\sqrt{\frac{E_{R}m}{\hbar^{2}}}$,
total energy $E=E_{r}+E_{R}$ and $J_{|l-\alpha|}$ is a Bessel function of the first kind. The box normalization of $\sqrt{k/L^3}$
is strictly only valid when $k>0$ and in the large $L$ limit. We
see that different sorts of anyons have different probabilities to be close to
one other (Fig~\ref{fig:wavefunction}). In particular, bosons are the only particles which can
be at the same point ($r=0$) since $J_{|l-\alpha|}(0)>0$ only if $l=\alpha=0$;
other anyons satisfy a hard-core condition for all $l$.  Spectral functions for creating localized excitations can be expected to show signs of
these differences, which reflect repulsive angular momentum barriers.

\paragraph*{Spectral function}

We consider the Lehmann representation\cite{mah00} of a zero temperature
spectral function $S(\vec{Q},\omega)$ associated to a spectroscopic measurement 

\begin{align}
S(\vec{Q},\omega) & =\sum_{|\psi_f\rangle}\left|\langle \psi_f|\hat{O}_{\vec{Q}}|\psi_i\rangle\right|^{2}\delta(\omega+\omega_{i}-\omega_{f})
\end{align}

\noindent Here, $|\psi_f\rangle$ are two anyon energy eigenstates and
$|\psi_i\rangle$ is the topologically ordered ground state. We will look at the behaviour 
close to a momentum $\vec{q}$ with quadratic dispersion. We assume the operator
$\hat{O}(\vec{R})$ creates a superposition
of two particle excitations whose center of mass is located at $\vec{R}$
and are separated from each other by a distance $a$, i.e., $\hat{O}(\vec{R})|\psi_i\rangle=\int d\phi|\vec{R},a,\phi\rangle$.
This is motivated from the creation of excitations in lattice models
with topological order such as the toric code where a local operator
such as $\sigma_{z}$ creates excitations on neighbouring sites. Breaking rotational invariance doesn't affect the final answer at low energies where s-wave eigenstates ($l=0$) dominate. In general, $\hat{O}(\vec{R})$ 
creates some local perturbation to the many-body ground state (like a certain spin texture) and the matrix element reflects the overlap of this state with two anyon many-body eigenstates. However, it can be shown that the resulting energy dependence near the threshold is insensitive to details (Appendix~\ref{appendix:universality}).
We set $\hbar=1$ henceforth.

\begin{align}
& S\left(\vec{q},\omega\right) \nonumber \\ = &c\sum_{|\psi_f\rangle}\left|\int d\vec{R}d\phi e^{i\vec{q}\cdot\vec{R}}\langle \psi_f|\vec{R},a,\phi\rangle\right|^{2}\delta(\omega+\dfrac{k^{2}}{m}-\dfrac{|\vec{K}|^{2}}{4m}-\Delta)\label{eq:str}\\
 =&cmJ_{\alpha}^{2}(a\sqrt{\Omega})\Theta\left(\Omega\right)\label{eq:str3}\\
 \approx & cm(a^{2}\Omega)^{\alpha}\Theta\left(\Omega\right)\label{eq:two_free}
\end{align}

\noindent where $\Omega=m(\omega-\Delta)-|\vec{q}|^{2}/4$, $\Delta$
is the energy gap to the threshold of two-particle excitations, and in the last step we have made a low-energy approximation. An infrared cutoff is
set to avoid spurious divergences and $c$ is an energy independent constant. 

Above the gap, bosons have
a sharp onset whereas fermions show a linear increase with energy
and semionic excitations ($\alpha=0.5$) show a characteristic square root dependence.
It is important to note that the difference arises due to the effect of statistics
on the matrix elements and is not an effect from the density of states. This can 
be seen easily for bosons and fermions, where the two-particle density of states is the same 
in the thermodynamic limit, since Pauli exclusion for fermions only excludes a one-dimensional line in the three-dimensional momentum space available to two bosons at fixed total momentum.

Although we have only considered the case of identical excitations so far, 
the results generalize to the case of two distinct particles interacting 
through mutual statistics.  There are a few minor differences, such as the presence of two distinct masses and the ability of (formerly)``bosonic'' angular momentum $l$ to assume odd values.

\paragraph*{Effect of interactions between anyons}

In general, the projection of the many-body problem onto the two-particle
subspace will induce interactions between the excitations, i.e., $V\left(r,\phi\right)\neq0$.
In parton constructions of gapped spin liquids, there is an emergent
gauge field which can mediate short-range interactions between the
excitations\cite{Wen2002}. In the absence of resonances, weak short-range interactions
will generically not affect the power law of the dynamic structure factor at low
energies. This can be seen from the fact that the anyon eigenstates are rigid at short 
distances (where interactions dominate) near the threshold and are only modified at large distances in interaction-free regions. This means that  $S\left(\vec{q},\omega\right)$ only changes by an overall energy-independent factor near the threshold (Appendix~\ref{appendix:universality}).

However, bosons do get a correction since the non-interacting point is fine-tuned. 
Put differently, since bosons lack a statistical repulsion
that prevents them from getting close to each other, they are susceptible
to short-range repulsive interactions. To quantify this, we obtain
$S\left(\vec{q},\omega\right)$ for a system with two 
excitations which are hard-core bosons, i.e., bosons which interact
with a hard-core potential $V(r)$ which is infinite for $r\leq b$
and zero everywhere else. The general form of the eigenstates is 
\begin{gather}
\Psi\left(R,r,\phi\right)=\exp(i\vec{K}\cdot\vec{R})\exp(il\phi)\left[A_{l}J_{l}\left(kr\right)+B_{l}N_{l}\left(kr\right)\right],\label{eq:hardCore_eigenstates}
\end{gather}
\\
where we may focus on $l=0$ since this has the most important contribution at low energies.

The effect of the potential can be incorporated into a phase-shift
defined by $\tan\delta_{0}=-B_{0}/A_{0}=J_{0}(kb)/N_{0}(kb)$. 
\noindent Normalization of the eigenstates yields $A_{0}=\cos\delta_{0}\sqrt{k/2L}$
where $L$ is the radius of the system.
We can now obtain $S\left(\vec{q},\omega\right)$ using  Eq.~(\ref{eq:hardCore_eigenstates})
\begin{align}
& S\left(\vec{q},\omega\right) \nonumber \\ = & \dfrac{cm}{1+\tan^{2}\delta_{0}}\left|J_{0}\left(a\sqrt{\Omega}\right)-\tan\delta_{0}N_{0}\left(a\sqrt{\Omega}\right)\right|^{2}\Theta\left(\Omega\right)\label{eq:hcb0}\\
\approx & \dfrac{cm\log^{2}\left(\dfrac{b}{a}\right)}{\left(\log\left[\dfrac{\Omega b^{2}}{4}\right]+2\gamma\right)^{2}+\pi^{2}}\Theta\left(\Omega\right)\label{eq:hcb}\\
\nonumber 
\end{align}

\noindent where $\Omega$ and $a$ are the same as in the free anyons
case, $\gamma$ is the Euler Mascheroni constant and we have made
a low-energy approximation in the last step. We see that the hard-core
interaction drastically changes the low-energy behaviour for bosons, and one can also show that
 any finite repulsive interaction produces a similar
effect. A corresponding analysis for semions and fermions shows that interactions do not affect
their low-energy behaviour, as expected. However, long-range
interactions, such as Coulomb interactions, can affect the low-energy
behaviour. 

\paragraph*{Numerics}

\begin{figure}
\includegraphics[width=1.1\columnwidth]{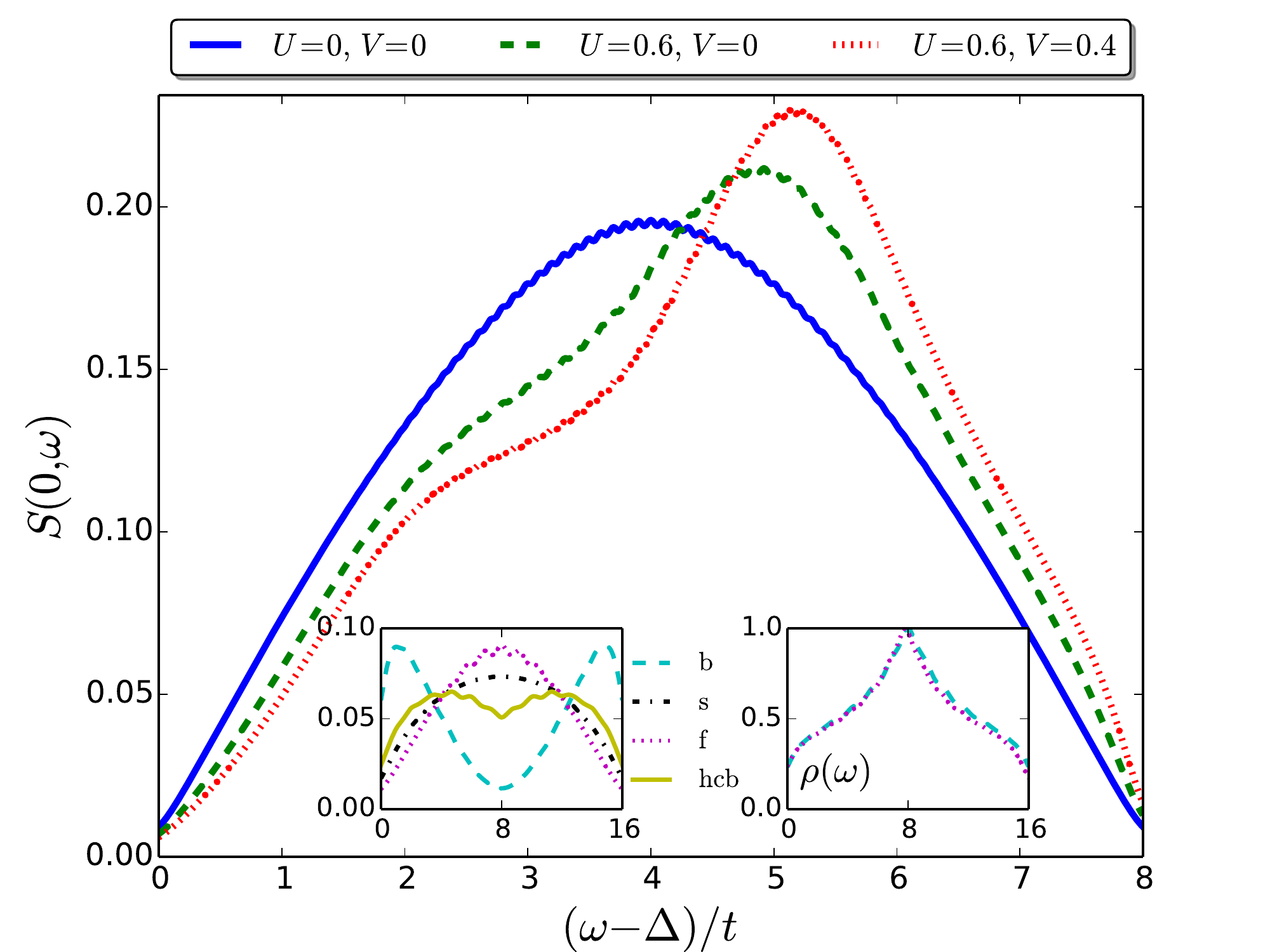}\protect\caption{Two particle spectral function from exact diagonalization (ED) on a $200\times200$ square lattice for fermions with nearest neighbor repulsion $U$ and next-nearest neighbor repulsion $V$. The high energy behavior is drastically affected by interactions, but the low-energy linear onset is unchanged. The on-set is not exactly zero because of the broadening involved in ED with $\epsilon$=0.1. The left inset compares $S\left(0,\omega\right)$ for non-interacting bosons (b), semions (s), fermions (f), and hard-core bosons (hcb) on a $20\times20$ lattice. The right inset shows the corresponding density of states for bosons and fermions. \label{fig:S(q,w)}}
\end{figure}

We can check some of the results numerically by considering an effective model of two anyons 
hopping on a square lattice which can be constructed from a many-body system by projecting into 
the two anyon subspace. We assume a local operator creates anyons on neighbouring sites of the 
lattice and perform exact diagonalization to obtain the corresponding spectral function.  
Fig.~\ref{fig:S(q,w)} shows $S\left(0,\omega\right)$ for fermions 
interacting with nearest neighbor repulsion $U$ and next-nearest 
neighbor repulsion $V$. Interactions drastically affect the high-energy 
behavior, but leave the low-energy linear onset unchanged as expected 
for fermionic excitations.

We can also obtain the spectral function for general anyons which are modeled as a 
charges living on the sites of the lattice which always move together with fluxes living on an adjacent
plaquette\cite{Einarsson90, *WenTorus1990, *Kallin93}. Since we use 
periodic boundary conditions, there are large gauge transformations of the vector
potential which result in doubling the Hilbert space for semions.
Fig.~\ref{fig:S(q,w)} (left inset) shows $S\left(0,\omega\right)$ for particles of various statistics
and the behavior at low energies is qualitatively similar to
the analytic predictions for two anyons in the continuum (Eq.~\ref{eq:two_free},
\ref{eq:hcb}). Bosons shows a decreasing behaviour which fits in
with the fact that their low-energy dependence is dominated by $J_{0}(x)$.
The square root energy dependence for semions means that they show
a more rapid increase initially as opposed to the linear onset for
fermions. As expected, the density of states (Fig.~\ref{fig:S(q,w)} right inset) is
the same for the different particles and does not 
the behaviour of the spectral function. 

\paragraph*{Three particles}

We now consider a system with three identical
fractionalized excitations ($n=3$) where we see non-trivial effects due to the braiding 
of three particles around each other. Unlike bosons and fermions,
a system of three anyons is no longer exactly solvable even without
interactions ($V\left(r,\phi\right)=0$), but the approximate low-energy
dependence of $S\left(\vec{q},\omega\right)$ can still be obtained
for certain anyons. After separating out the center of mass motion,
the resulting system\cite{Wu1984, *Mashkevich1995} can be described
in terms of hyper-spherical coordinates consisting of a radial co-ordinate
$\rho$ and three angular co-ordinates $\theta,\phi,\psi$. This again
reduces the relative system to an effectively one-dimensional problem.

The crucial part of obtaining the low-energy dependence of the spectral function 
lies in the fact that the radial solution is determined by the eigenvalue of the 
angular equation which can be obtained by relating it to the ground state energy of 
the system in a harmonic oscillator potential which is known analytically (Appendix~\ref{appendix:3anyon}).
We can thus obtain the eigenstates of the free problem
with zero total angular momentum, which are the dominant
contribution to $S\left(\vec{q},\omega\right)$ at low energies as in the
two particle case. 

In particular, for the interesting case
of three fractionalized particles with $\alpha=\frac{1}{3}$, the
eigenstates are $\psi\left(\rho,\theta,\phi,\psi\right)=\frac{J_{2}\left(\sqrt{2E}\rho\right)}{\rho}g\left(\theta,\phi,\psi\right)$
where $E$ is the energy of the relative motion and $g$ is a function
which is independent of the energy. This leads to a threshold behaviour of $S\left(\vec{q},\omega\right)$  which increases as $\left(\omega-\Delta\right)^{2}$. 

\paragraph*{Applications}
Our results (eq. (\ref{eq:two_free}) and (\ref{eq:hcb})) could be used to study the statistics of excitations in a spin liquid or a fractional Chern insulator by measuring the scattering cross-section as a function of energy close to the threshold.  At low energies, the excitations would be described by Schrodinger's equation (as assumed above) because the dispersion can generically be expanded to quadratic order around the minimum. 
We require that the mean free path of the excitations is much longer than their wavelength. If the mean free path is long enough, one can measure the structure factor slightly above the threshold, so that the disorder can be neglected, but not so far above the threshold that the universal behaviour is lost.

Consider an experiment on a chiral spin liquid material.  A dramatic consequence of these results is the prediction that the threshold behaviour one would have a square root onset with energy. Chiral spin liquids have one type of topological excitation, which are semions. A single semion is not accessible from the ground state by local processes, such as inelastic scattering by neutrons due to the conservation law that the total topological charge must be trivial. However, a neutron could excite two semions because semions are their own antiparticles, which then leads to the square-root onset.

In the case of materials in the toric code universality class (i.e. $Z_2$ spin liquids), one could also measure anyonic statistics. The possible topological charges of excitations are $m$ and $e$ (which are separately bosons but have a mutual phase of $-1$) and a composite  $em$ which is a fermion.

 The most interesting thresholds (consistent with the topological conservation law which requires $e$'s and $m$'s to be created in pairs) are the production of an $(me)-(me)$ pair, which would show 
 a linear onset because $me$ is a fermion, and a triple, $e-m-(me)$. (The pairs $m-m$ and $e-e$ are both
 mutual bosons, so their structure factor would have the same form as that possessed by local excitations such as two triplon excitations in a VBS state.)
 There are two difficulties in this case; first, there may not be a stable particle of type $me$ (in a spin liquid, the stable spinon might have the type either $m$ or $me$).
 Second, the excitations also carry spin; therefore the pair of $me$ particles can either be in a triplet channel or a singlet channel (which would have the same behavior as a pair of bosons). This can be avoided by applying a weak magnetic field to favor parallel spins.
 
It is also appropriate to mention that the results do not apply, in a straightforward way, within the quantum Hall complex.  This is primarily because the quasiparticles in those states are generally electrically charged.  The Coulomb force then formally dominates any effective statistical force at large distances, and it will be quantitatively significant even if it is weakly screened.  In addition, the motion of charged quasiparticles is inhibited by the background magnetic field. The presence of the magnetic field causes the appearence of discrete responses in spectroscopy instead of a threshold due to the formation of Landau levels by the quasiparticles. These problems are avoided in fractional chern insulators where we can expect to see the universal threshold behaviour.



There are also measurable signatures in tunneling spectroscopy 
which can involve tunneling of neutral particles or cases where 
the Coulomb interaction between the excitations is screened. We expect a different exponent governing tunneling since the particles do not have a well-defined momentum any more. For the simplest case where the tunneling tip resembles a Fermi liquid, we find that the tunneling current goes as $(\omega - \Delta)^{1+\alpha}$. 

Several recent works\cite{Qi2009, Kamfor2014,Punk2014, Shubhayu2015} contain calculations of dynamic structure
factors in topologically ordered phases of spin systems. We now consider three prominent
examples. 

Qi et al\cite{Qi2009} describe a bosonic $\mathbb{Z}_{2}$ spin liquid on the 
triangular lattice where the low-energy behaviour near the bottom of the band is obtained through 
a large-N analysis of a sigma model. There is a constant onset 
above the gap for non-interacting bosonic spinons which changes to an inverse-log behaviour 
on adding interactions, exactly as expected. For the case of the $Q_{1}=Q_{2}$
spin liquid on the kagome lattice\cite{Sachdev1992}, the spinons have
a quadratic dispersion around the high symmetry $\mathrm{M}$ point and the structure
factor\cite{Punk2014} there shows a sharp onset above the gap, as expected for bosonic
excitations. Finally, Kamfor et al\cite{Kamfor2014,Kamforthesis} study the toric code\cite{Kitaev2003}
in a weak magnetic field where the relevant operator creates electric defects which have a quadratic dispersion at the $\Gamma$ point and there is a rapid increase initially and a minimum
in the middle of the spectrum, as expected for hard-core bosons with weak 
interactions \footnote{The minimum in the middle of the spectrum can be seen clearly when a rotationally symmetric operator is evaluated, as in \cite{Kamforthesis}}

\paragraph*{Discussion}

We have shown that the production rates for
fractionalized excitations (as could be inferred from the dynamic correlation functions as measured in neutron scattering or tunneling experiments) contain 
signatures of anyons. They follow a power law characteristic of their exchange and mutual statistics.
This may also be accessible in systems of ultra-cold atoms in optical lattices following recent ideas of how to access their spectral functions\cite{Goldman2012, *Yao2013}.    

We thank Collin Broholm, Curt von Keyserlingk, Chris Laumann, Roderich Moessner, Shivaji Sondhi and Michael Zaletel for useful discussions. This work was partly supported by the DFG grant SFB 1143 and FOR 1807 grant P01370/2-1. FW's work is supported by the U.S. Department of Energy under grant Contract Number  DE-SC0012567.

\bibliographystyle{apsrev4-1}
\bibliography{refs}

\appendix
\section{Universality near the threshold \label{appendix:universality}}
We give here an explanation of why the spectral function shows universal behaviour near the 
threshold away from resonances which is independent of arbitrary short-range interactions and also of the exact form and size of the local operator creating the excitations. We illustrate it by focusing on the case of two excitations (i.e., we consider energies close to the two-particle threshold and far from the next threshold).

Consider the matrix element $M = \left|\langle \psi_f|\hat{O}_{\vec{Q}}|\psi_i\rangle\right|^{2}$ where $|\psi_i\rangle$ 
is the ground state and $|\psi_f\rangle$ is an eigenstate with energy $\omega - \Delta$. 
Without loss of generality, let us consider the case with $\vec{Q}=0$. In real space, we have 
$M =  \left|\int d\vec{R} \langle \psi_f|\hat{O}(\vec{R})|\psi_i\rangle\right|^{2}$. Since we are close to the two-particle threshold, we can consider only two-particle eigenstates and do an approximate expansion of $\hat{O}(\vec{R})|\psi_i\rangle \sim \int dr d\phi \chi(\vec{R}, r, \phi) |\vec{R}, r, \phi\rangle$ for some coefficients $\chi$ (the notation $\vec{R}, r, \phi$ is the same as in the main text). As the operator $\hat{O}$ has a finite support, we get that $M  \sim \left| \int d\vec{R} dr d\phi \chi(\vec{R}, r, \phi) \Psi^{*}(\vec{R}, r, \phi) \right|^2$ where the integral over $r$ only runs up to some finite $r=r_o$. Now, assume that we separate the space into an internal region $r<r_s$ where short-range interactions dominate and an external region where the particles are  described by their free wavefunctions. If $r_s<r_o$, then we can use the same description as in the main text. 

\begin{figure}
\includegraphics[width=0.6\columnwidth]{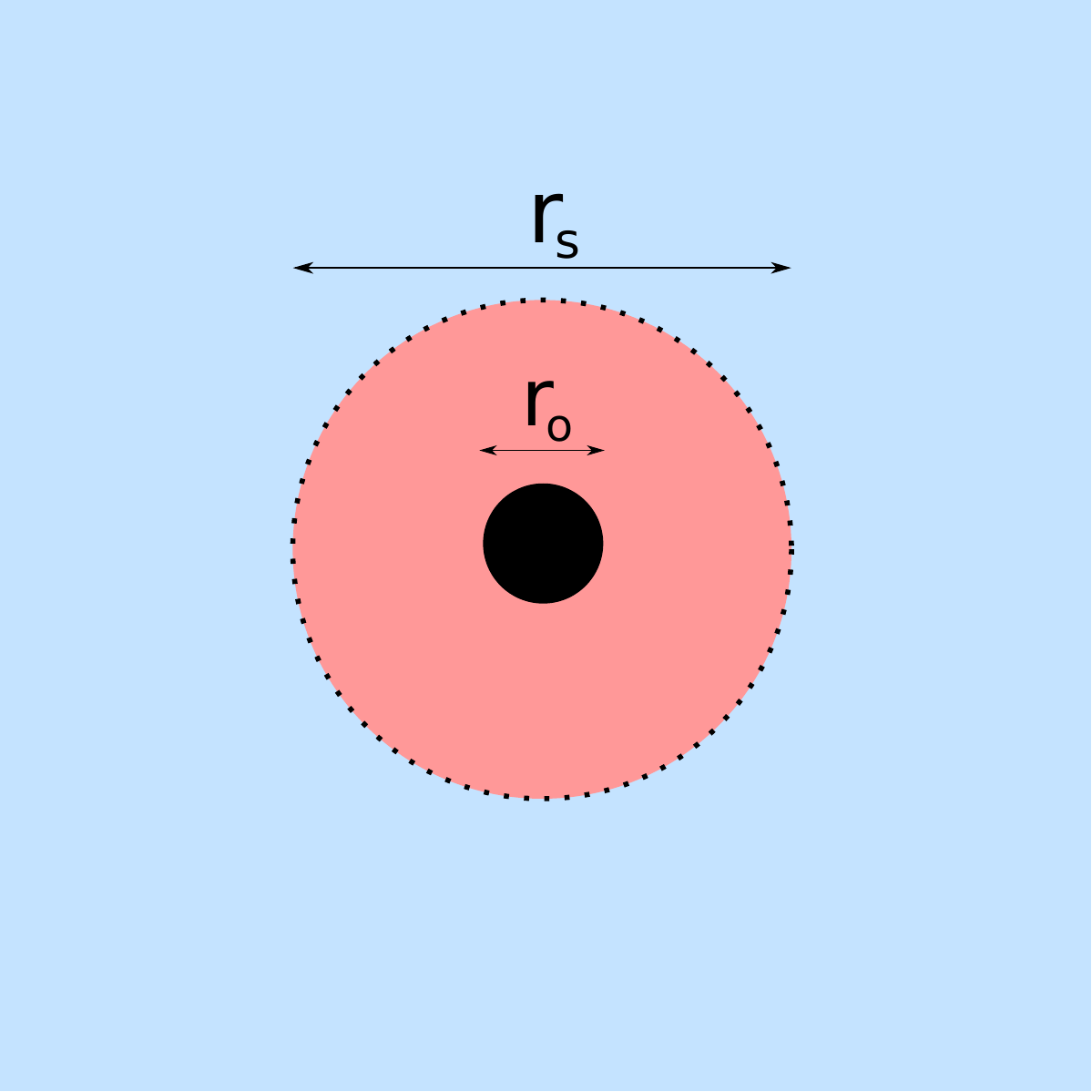}\protect\caption{The system can be divided into an internal region (dotted red circle with radius $r_s$) where short-range interactions dominate and an external region where the wavefunctions are described by those of free anyons. Although the operator $\hat{O}$ acts deep inside the internal region (small black circle with radius $r_o$), the wavefunction inside the internal region is rigid, i.e., it does not change with energy near the threshold and the matrix element can be related by a constant to the wavefunction at the boundary of the internal and external region.  \label{fig:universality}}
\end{figure}

Let us now consider the case $r_s>r_o$ (Fig~\ref{fig:universality}). Naively, it can seem that the matrix element might be dominated by the form of the short-range potential since we are probing the creation of particles in regions with strong short-range interactions. However, we note that the eigenvalue problem inside the internal region is dictated completely by the short-range potential (we can neglect the $k^2$ term in Schrodinger's equation near the threshold). This means that the eigenstates for the internal region do not vary much with energy and are "rigid" near the threshold. Now, we know that $\Psi(r=r^{-}_s) = \Psi(r=r^{+}_s)$. Since the internal eigenstates are rigid, this implies that $\Psi(r<r^{-}_s) = c \Psi(r=r^{+}_s)$ where $c$ is some function independent of energy. This means that we have $M \sim \left| \int d\vec{R} dr d\phi \chi(\vec{R}, r, \phi) \Psi^{*}(\vec{R}, r^{+}_s, \phi) c \right|^2$. The only energy dependence term in $M$ is the free-particle wavefunction at the boundary. Thus, we get the same answer for the spectral function near the threshold as in the main text up to an overall energy independent constant.

\section{Three anyon problem \label{appendix:3anyon}}

The three particle problem can be simplified by first separating into
the center of mass motion and the relative motion\cite{Wu1984,Mashkevich1995}.
The relative motion problem is described by hyper-spherical coordinates
$0\leq\rho\leq\infty$, $\frac{-\pi}{4}\leq\theta\leq\frac{\pi}{4}$,
$\frac{-\pi}{2}\leq\phi\leq\frac{\pi}{2}$, $0\leq\psi\leq2\pi$ and
the corresponding Hamiltonian for particles with statistics parameter
$0\leq\alpha\leq1$ is
\begin{equation*}
H_{r}=H_{0}^{rad}+\dfrac{1}{m\rho^{2}}\left(H_{0}^{ang}+\alpha H_{1}+\alpha^{2}H_{2}\right)
\end{equation*}
\noindent where
\begin{equation*}
H_{0}^{rad}=\dfrac{1}{2m}\left(-\dfrac{\partial^{2}}{\partial\rho^{2}}-\dfrac{3}{\rho}\dfrac{\partial}{\partial\rho}\right)
\end{equation*}
\noindent and $H_{0}^{ang}$ is the laplacian on the 3-dimensional
sphere, and $H_{1}$ and $H_{2}$ are functions of $\theta,\phi,\psi$
which encode the effects of statistics since they couple to $\alpha$.
This leads to two eigenvalue problems
\begin{gather*}
\begin{aligned}\left(H_{0}^{ang}+H_{1}+H_{2}\right) & =\lambda\psi\\
\left(H_{0}^{rad}+\dfrac{\lambda}{m\rho^{2}}\right)R(\rho) & =ER(\rho)
\end{aligned}
\end{gather*}
\noindent The solution of the radial part is given by 
\begin{equation*}
R(\rho)=\dfrac{c}{\rho}{\displaystyle J_{\sqrt{1+\lambda}}\left(\sqrt{2E}\rho\right)}
\end{equation*}
\noindent where $c$ is a constant. We can obtain $\lambda$ by noticing
that the angular eigenvalue equation is independent of any potential
$V(\rho)$. In particular, adding a harmonic oscillator potential
$m^{2}\omega^{2}\rho^{2}$ does not change the equation. This is useful
because the eigenvalue $\lambda$ is related to the energy eigenvalue
of the harmonic oscillator solution for three anyons. In particular,
the ground state energy of the harmonic oscillator\textcolor{black}{\cite{Wu1984, *Mashkevich1995}}
is $E_{0}=(\sqrt{\lambda+1}+1)\omega$. For particles with $\alpha<0.71$,
the exact ground state energy $E_{0}$ is known to be $(2+3\alpha)\omega$.
Hence, for particles with $\alpha=\frac{1}{3}$, we get $\lambda=3$.
Since the angular eigenstates do not depend on the energy, they will
only add an overall form factor to the spectral function $S\left(\vec{q},\omega\right)$.

\end{document}